\documentclass[10pt,letterpaper]{article}
\usepackage[top=0.85in,left=1.4in,footskip=0.75in]{geometry}

\usepackage{amsmath,amssymb}

\usepackage{changepage}

\usepackage[utf8x]{inputenc}

\usepackage{textcomp,marvosym}

\usepackage{cite}

\usepackage{nameref,hyperref}

\usepackage{microtype}
\DisableLigatures[f]{encoding = *, family = * }

\usepackage[table]{xcolor}

\usepackage{adjustbox}

\usepackage{array}

\usepackage{amssymb}
\usepackage{graphicx}
\usepackage{subcaption}
\usepackage{caption}

\usepackage{setspace}
\newcolumntype{+}{!{\vrule width 2pt}}

\newlength\savedwidth

\usepackage[aboveskip=1pt,labelfont=bf,labelsep=period,justification=raggedright,singlelinecheck=off]{caption}

\makeatletter
\renewcommand{\@biblabel}[1]{\quad#1.}
\makeatother

\usepackage{lastpage,fancyhdr,graphicx}
\usepackage{epstopdf}
\fancyheadoffset[L]{2.25in}
\fancyfootoffset[L]{2.25in}
\usepackage{xcolor}

\usepackage[normalem]{ulem}

\begin{document}
\vspace*{0.2in}

\begin{flushleft}
{\Large
\textbf\newline{Application of Machine Learning to Predict the Risk of Alzheimer's Disease: An Accurate and Practical Solution for Early Diagnostics} 
}
\newline
\\
Courtney Cochrane\textsuperscript{1},
David Castineira\textsuperscript{1*},
Nisreen Shiban\textsuperscript{1},
Pavlos Protopapas\textsuperscript{1},

\bigskip
\textbf{1}  Institute for Applied Computational Science, Harvard John A. Paulson School of Engineering and Applied Sciences, Cambridge, MA, US

\bigskip

%
%





* davidcastineira@outlook.com

\end{flushleft}
\section*{Abstract}
Alzheimer's Disease (AD) ravages the cognitive ability of more than 5 million Americans and creates an enormous strain on the health care system. This paper proposes a machine learning predictive model for AD development without medical imaging and with fewer clinical visits and tests, in hopes of earlier and cheaper diagnoses. That earlier diagnoses could be critical in the effectiveness of any drug or medical treatment to cure this disease. Our model is trained and validated using demographic, biomarker and cognitive test data from two prominent research studies: Alzheimer’s Disease Neuroimaging Initiative (ADNI) and Australian Imaging, Biomarker \& Lifestyle Flagship Study of Aging (AIBL). We systematically explore different machine learning models, pre-processing methods and feature selection techniques. The most performant model demonstrates greater than 90\% accuracy and recall in predicting AD, and the results generalize across sub-studies of ADNI and to the independent AIBL study.  We also demonstrate that these results are robust to reducing the number of clinical visits or tests per visit. Using a meta-classification algorithm and longitudinal data analysis we are able to produce a "lean" diagnostic protocol with only 3 tests and 4 clinical visits that can predict Alzheimer's development with 87\% accuracy and 79\% recall. This novel work can be adapted into a practical early diagnostic tool for predicting the development of Alzheimer's that maximizes accuracy while minimizing the number of necessary diagnostic tests and clinical visits.

\section*{Author summary}
The main goal of this paper is to propose a machine learning solution for the problem of predicting the risk of developing Alzheimer's Disease (AD). This is achieved by systematically analyzing medical records from two of the longest longitudinal studies of AD, ADNI and AIBL. We analyze different machine learning algorithms as well as feature selection methods and preprocessing techniques. Our proposed solution encompasses a diagnostic protocol for early testing of AD that has high accuracy and recall while also minimizing the number of diagnostic tests the patient is subjected to and eliminating the need for costly imaging data. This renders our solution both accurate as well as practical for an early detection program of AD.


\section{Introduction}


Alzheimer's Disease (AD) is the most common cause of dementia, a group of brain disorders that cause the loss of intellectual and social skills. AD manifests as a progressive, degenerative disorder that attacks the brain's nerve cells, or neurons, resulting in loss of memory, thinking and language skills, as well as behavioral changes \cite{AFA}. Currently AD is an irreversible process with no cure. The personal, social and economic impact of AD is profound: In the United States, more than 5 million people aged 65 or over are suffering from Alzheimer's disease and the estimated national cost of patient care for Alzheimer's and other dementias was \$236 billion in 2016 \cite{ALZ}. Further, AD is the sixth leading cause of death in the US (third for older people) \cite{ALZ2}. 

Although Alzheimer's  was first identified more than a century ago, effective treatments have proved elusive. Drug and non-drug treatments can help alleviate some cognitive and behavioral symptoms of AD, but there is still no cure. Researchers continue to work on developing treatments that can reverse disease progression and improve the quality of life for people with Alzheimer's. One of the critical challenges for dealing with AD is the lack of understanding about the neurodegenerative process associated with this disease. There are currently two widely-believed, competing hypotheses:
\begin{enumerate}
\item[1)] The amyloid hypothesis: One prime suspect for AD is a microscopic brain protein fragment called beta-amyloid. This protein is a sticky compound that accumulates in the brain, disrupting communication between brain cells and eventually killing them. Some researchers believe that flaws in the processes governing production, accumulation or disposal of beta-amyloid are the primary cause of Alzheimer's \cite{ALZbeta}.
\item[2)] The tau hypothesis: The accumulation of the tau protein is thought to be a major player in the development of Alzheimer's disease. In particular, the tau hypothesis asserts that the formation of neurofibrillary tangles (insoluble twisted fibers that are formed inside the cells) causes the development of AD \cite{Mohandas}.
\end{enumerate} 

Given the huge social and economic impact of any potential treatment for AD, several companies are actively researching this field. Recent research has created optimism that a treatment for AD is close to fruition \cite{Fillit}. Regardless of the treatment, one critical aspect for the practical deployment of any potential AD drug is the ability for this drug to be used widely and preventively \cite{MIT}. Thus, this work aims to provides a model, using two well-known studies of Alzheimer's disease (ADNI and AIBL), that is both suitable for early detection and practically applicable in clinical settings. To achieve this aim, we strategically evaluate our model's performance with the smallest (and least expensive) feature subsets so that our model can be used for early screening, before any symptoms appear. Our model utilizes features including demographics, biomarkers, and cognitive tests. Due to the prohibitive expense of medical imaging data (e.g., MRI and PET scans) as an early detection test, we remove medical imaging as a possible feature. Our machine learning approach uses metrics derived from longitudinal data analysis, and our analysis evaluates optimal feature selection techniques, data imputation methods, and classification algorithms. Ultimately our goal is to provide a cost-effective pre-screening test battery for Alzheimer's disease.\\

\subsection{Literature Review}
We explored existing literature regarding Alzheimer's prediction as well as longitudinal data handling. Prediction of Alzheimer's is a popular area of research with researchers applying a plethora of supervised learning techniques to the problem. We also found existing literature that worked with the same main data set that we use in our study, ADNI. In general we find that Support Vector Machines are the most popular machine learning technique applied to this problem \cite{Kloppel} \cite{Orimaye}. However, recently Neural networks have gained popularity \cite{Hosseini}, and novel approaches have been attempted, including using Natural Language Processing to find linguistic deficits \cite{Orimaye}. The biggest difference between most studies predicting Alzheimer's and our work lies in our exclusion of medical imaging as a feature for the model. Our goal was to produce a model that could aid in early detection, and the cost of medical imaging is a deterrent for people who are unsure about getting tested for the disease. Therefore, we excluded medical imaging (PET and MRI) from our analysis. The literature that uses medical imaging has produced accuracies of 80\% \cite{Korolev} and recalls of 85\% \cite{Devanand} in predicting conversion to Alzheimer's. The few models that do not include medical imaging data, instead using cognitive tests and demographics, have accuracies of less than 85\% \cite{Datta}. Ultimately, this is a ripe general area for research, but there are a dearth of studies considering prediction without the use of medical data. \\

We also conducted a literature review for longitudinal data analysis, which is very relevant for the type of data typically associated with AD clinical studies. In our case, the longitudinal nature of the AD data results from the observation of subjects (patients) over time during sequential clinical visits. The difficulty in dealing with this data stems from the inconsistent number of observations across patients and the potentially correlated data within patients. The literature for longitudinal data analysis is extensive \cite{Locascio} \cite{vanBelle}. 

Methods based on summary metrics or statistics \cite{Fitzmaurice} have been broadly used, where temporal measurements are summarized into key statistical descriptors (e.g., mean, mode, area-under the curve, etc). Another common solution to handle longitudinal data is to fit this data using some type of regression model. Regression models permit inference regarding the dynamic response over time and how this evolution varies with patient characteristics such as treatment assignment or other demographic factors. However, standard regression methods assume that all observations are independent, and this may produce invalid standard errors if the assumptions are not valid. For this reason, advanced regression methods have been proposed to overcome some of these limitations such as Random-Coefficient Models \cite{Rutter} and General Regression Methods \cite{Fitzmaurice2}. Some of these models are very flexible in allowing for imbalanced data, missing values, differing number of time points from subject to subject, and unequal spacing of time point intervals within and across subjects. In addition, recent work has been done applying machine learning techniques such as Neural Networks \cite{Tandon} and Support Vector Classifiers \cite{Chen} to longitudinal data. For the work presented here, summary metrics have been shown to provide great results in generating features for predictive modeling, with the added benefit of generating more parsimonious models. 

\section{Materials and methods}
\subsection{Data}
For this study we have utilized two existing repositories for AD studies: ADNI \cite{ADNI} and AIBL \cite{AIBL}: \\

The ADNI (Alzheimer's Disease Neuroimaging Initiative) dataset is an ongoing, longitudinal multi-study that has been carried out since 2004. It has acquired data and specimens from 1,700 participants at 60 clinical sites around Canada and the United States. The study enrolls selected populations for future treatment, and the subjects include AD patients, mild cognitive impairment subjects, and an elderly control. The successes of the ADNI database includes developing standardized methods, improving trial efficiency, and creating an infrastructure for sharing raw and processed data without embargo. The initiative is supported by $\$67$ million in private and public sector donations. The initial phase of the study is known as ADNI1. In 2009, the second phase, ADNIGO, was started containing 200 participants with Early Mild Cognitive Impairment. In 2011, the third phase, ADNI2, began with 150 participants with Late Mild Cognitive Impairment.

The AIBL (Australian Imaging, Biomarker \& Lifestyle Flagship Study of Aging) dataset contains data from a 4.5 year longitudinal study of cognition which started in 2006. It is a large scale cohort study containing 1,112 participants and conducted at two centers, Perth and Melbourne, in Australia. The study focuses on early detection, specifically in terms of lifestyle interventions. The AIBL data contains 211 AD patients, 133 MCI patients, and 768 healthy volunteers and follows the ADNI1 protocols for data collection. The available data includes clinical and cognitive data, image data (extracted from MRI and PET data), biomarker data including blood, genotype, and ApoE, and dietary and lifestyle data. These latter assessments examine participant's diet, exercise patterns, body composition, and sleep habits.\cite{AIBL}. 

More detailed clinical description of the ADNI and AIBL cohorts have been previously published in \cite{ADNI_clinical} and \cite{AIBL} respectively. It is worth noticing that AIBL and ADNI have many of the same goals and are designed to identify the biomarkers, cognitive characteristics, and health and lifestyle factors that impact AD. 

For this study we used 94 predictors that were reported in the ADNIMERGE table (a special dataset that merges key ADNI tables). These predictors provided specific information about patient demographics, genetics, blood biomarkers and cognitive tests from participants in different longitudinal multi-center studies.

\subsection*{Model Procedure}

We experiment with multiple different pre-processing techniques, feature selection methods, and machine learning models. The specific options we experimented with are delineated below. 

\begin{itemize}
\item \textit{Data-Preprocesing}: Our data pre-processing includes generation of the labels, conversion of categorical variables, longitudinal data handling and imputation. First, we generated labels for our classification problem by merging features across the different data files in the ADNI dataset. We then excluded $14$ subjects who were diagnosed as AD but then were diagnosed as either cognitively normal (CN) or mild cognitive impairment (MCI) in future years (there is currently no way to reverse AD so this indicates a mistake in the data). 

We next one-hot encoded all categorical variables and performed feature engineering. For all longitudinal features, we computed a series of summary metrics (mean, standard deviation, absolute changes and time intervals) for each patient that captured their temporal evolution along multiple clinical visits. Finally, we split the data into a training and test set and performed imputation. We investigated imputation by mean or mode (mean for numerical columns and mode for categorical columns), and k-Nearest Neighbors imputation.\\   

\item \textit{Feature Reduction and Selection}:
Our data is high-dimensional, so we experimented with two different feature reduction techniques: Singular Value Decomposition and Affinity Propagation. Singular Value Decomposition, or SVD \cite{Wall} operates by combining information from several (likely) correlated vectors, and forming basis vectors which explain most of the variance in the data and are guaranteed to be orthogonal in higher dimensional space. SVD and PCA (Principal Component Analysis, a very popular dimensionality reduction technique) are closely related. On the other hand Affinity Propagation, or AP \cite{Frey07clusteringby} is a relatively new clustering algorithm based on the concept of "message passing" between data points. Once we obtain clusters of features then we can compute the so-called exemplars (features that are good representatives of themselves and some other features). This approach provides an elegant feature selection technique. Notice that AP does not require the number of clusters to be determined or estimated before running the algorithm (and this is in contrast to other clustering techniques such as k-means), although a user of this technique must sill define some hyperparameters (e.g., preferences) that affect the resulting number of clusters. \\

\item \textit{Supervised Learning}:
The supervised learning module performs five-fold cross validation and grid search over the hyperparameters and model selection specified in the pipeline. The models implemented in our pipeline were Random Forest, Logistic Regression, k-Nearest Neighbors, Support Vector Machines (SVM), Multi-layer Perceptron (MLP), AdaBoost, Linear SVM, Gradient Boosting, and Decision Trees. More details for these methods can easily be found in machine learning literature \cite{Bishop} \cite{Duda}. Using the parameters that maximize recall on the validation set, the model predicts and outputs the labels for the test set.

\item \textit{Model Evaluation}:
Finally, the predictions of our model are evaluated against the true labels using the following metrics: confusion matrix, accuracy, recall, precision, f1 score, and ROC curve. In this work, we focus on optimizing model accuracy (percentage of patients correctly labeled by our model) and recall (percentage of patients who develop AD that are correctly identified) \cite{AccuracyRecall}. In medical settings, like predicting AD development, it is often crucial to minimize false negatives, and therefore we try to optimize the recall of our models. However, our pipeline automatically computes the full suite of metrics for potential use in further exploration.
\end{itemize}

We evaluated every possible combination of model parameters (imputation, feature selection and model type) in order to quantify their predictive power. For each possible model, we assess the performance on fifty different random training and test set splits where the ratio of training to test data is 2:1. For these experiments, we use the entire ADNI dataset. We carry out five-fold cross-validation to pick the top four model/ hyperparameter combinations that maximize recall on the validation set, and record metrics across one hundred random splits of the train and test set to prove consistency of our results.\\ 

\subsection{Evaluation of Model's Promise for Early Detection of AD}
We also investigated whether our model can generalize, how much longitudinal data we need before we can make accurate predictions, and whether we can produce a "lean" model that minimizes time/money cost while maintaining high accuracy and recall.

\subsubsection{Model Generalization}
We also conducted a series of analyses in order to determine whether our model can generalize across different study protocols. We first explore the robustness of our top two models on the different sub-studies of ADNI. We train and test the model separately on ADNI1, ADNI2, and ADNIGO and record our four performance metrics. Next, we train a model on the ADNI1 dataset and then test separately on the ADNI2 dataset and the ADNIGO dataset. Likewise, we try all other pairs of sub-studies for the training and test sets. We also analyzed our model's performance on the AIBL dataset. Note that AIBL represents a completely different repository of patients to ADNI (i.e., different patients following different protocols), which gives us an excellent opportunity to validate our data-driven solutions for AD prediction. For this particular study we considered the handful of features that are common to both AIBL and ADNI repositories: Age, Gender, APOE4 (a genetic test) and MMSE (a cognitive test). 

\subsection*{Longitudinal Data Analysis}
As described earlier, one of our goals is to propose a practical data-driven solution for AD predictions that can be used in clinical settings. For this purpose, it is important to understand how the longitudinal dimension of the data (e.g., number of visits) affect the performance of our predictions. To evaluate this, we trained our model assuming that the full history of the patients in the training set was available. Then we tried to predict the label for the test set patients with restricted information from a limited number of visits. This analysis aims to evaluate the number of medical visits necessary for obtaining a given performance in predicting AD. The relationship between the number of visits and total time of study (from baseline to last visit) was also considered in this analysis. 

\subsection*{Cost Analysis}
To evaluate whether we can limit the cost for a patient while still providing an accurate prediction, we take two approaches. First, we research the time that patients spend on each test and then plot the features of our model against the time needed to obtain those features, with the features ordered by feature importance (as determined by the Random Forest algorithm). \\

Secondly, we utilize a meta-classification algorithm to produce models that have high accuracies and recalls while minimizing the testing time for the patient. We follow the approach of \cite{Pavlos} and build a meta-model, a Decision Tree, that balances accuracy of prediction with time cost (for the patient). First we generate the meta-classification dataset. We group our features into twelve different categories: demographics (e.g. age, gender, education), APOE4 genetic marker data, information about the number of years since the baseline diagnosis, and nine different cognitive test features. We first produce the power set of the original twelve features. Then we take all sets in the power set that have size 1, 2 or 3 (a total of $298$ sets to be considered). For each of these sets of features, we train a Random Forest model on the given features. The labels predicted by each of these $298$ models become a feature in our meta-classification dataset. The meta-classification model is a decision tree which splits based on the algorithm designed by \cite{Pavlos}. Instead of choosing a node to split on based on information gain, we choose the node, $M_i$, that has the maximal $\frac{Information\_Gain(M_i)}{Expected\_Cost(M_i)}$ where Information Gain is Shannon Information Gain and Expected Cost is defined as $Expected\_Cost(M_i) = P_L(M_i) Cost(M_i) + (1 - P_L (M_i)) [\sum_{v \in C_{M_i}} \sum_{j = i+1}^m P_L(M_j|M_i = v) Cost(M_j|M_i = v)]$ \cite{Pavlos} . Ultimately, this splitting criterion balances information gain with the cost of using the given model. We use this algorithm to create a decision tree that balances feature cost and information gain.

\section*{Results}

\subsection*{Comparison of Models}
\begin{figure}
\centering
\includegraphics[width=0.7\textwidth]{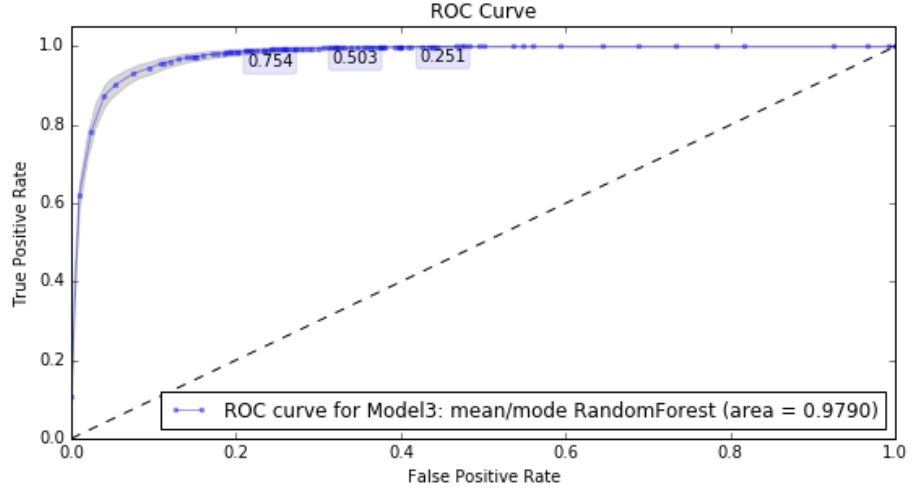}
\caption{\label{fig:roc} ROC Curve for Random Forest with mean/mode imputation (mean for numerical features and mode for categorical), no feature selection, and 1,000 trees. Small labels over the curve show the different thresholds used to make the class predictions.}
\end{figure}

For all the model parameter combinations, we determine that the best two models (with the highest recall on the validation set) are:~1. Random Forest with mean/mode imputation, no feature selection, and 1,000 trees and 2. Random Forest with k-Nearest Neighbors imputation, no feature selection, and 1,200 trees. Table~\ref{tab:ModelComparisonTable} shows the performance in terms of accuracy, recall, precision and F1 score. These scores are averaged across one hundred different random training and test set splits. Both our top models glean an average accuracy greater than 92\% and an average recall greater than 91\%, with the Model 1 having higher accuracy and Model 2 having higher recall. We produce an ROC curve for Model 1 which demonstrates the strong model performance (Figure \ref{fig:roc}).

\begin{figure}
\centering
\includegraphics[width=0.7\textwidth]{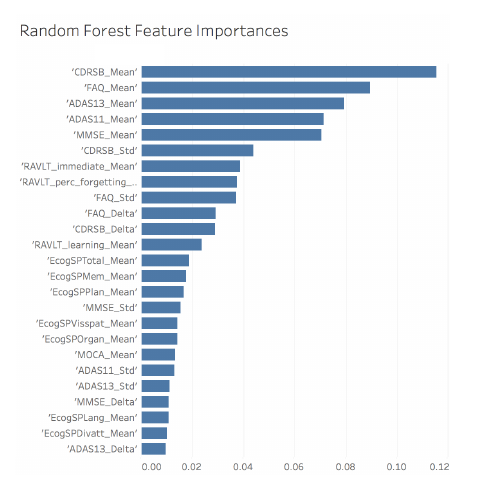}
\caption{\label{fig:featimp} Feature Importances for Top 25 Features in Model 1}
\end{figure}

Further we identify the features that are most important for Model 1 (Figure \ref{fig:featimp}). As shown, the top five most important features are related to longitudinal metrics for CDRSB, FAQ, ADAS11, ADAS13, and MMSE, which are all cognitive tests.  \\

\begin{table}[htb]
\centering
\caption{Comparison of Metrics for Top 2 Models}
\label{tab:ModelComparisonTable}
\begin{tabular}{|c|c|c|c|c|}
\hline
                 & \textbf{Accuracy} & \textbf{Recall} & \textbf{Precision} & \textbf{F1 Score} \\ \hline
\textbf{Model 1} & 92.51             & 91.76           & 89.33              & 90.48             \\ \hline
\textbf{Model 2} & 92.44             & 92.04           & 88.84              & 90.35             \\ \hline
\end{tabular}
\end{table}

\subsection*{Model Performance on ADNI sub-studies}

We test Model 1 and Model 2 when trained/tested on ADNI1, ADNI2, and ADNIGO separately. The models based on ADNI1 data and ADNI2 data have similar accuracies and recalls compared to the models trained on the full ADNI dataset. The models trained on solely ADNIGO data has a high accuracy, but low recall with high result variance. This is most likely a consequence of the different demographics of the sub-studies: only 10\% of the ADNIGO patients developed AD versus 30\% and 50\% patients in ADNI2 and ADNI1, respectively. Further, ADNIGO only has $129$ patients versus $789$ in ADNI2 and $819$ in ADNI1, making the training sample much smaller for ADNIGO. \\

Finally, for each pair of sub-studies, we train on one study and test on the other to determine whether our model generalizes across the different sub-studies. The different sub-studies have slightly different protocols and patient demographics, so we explore whether the results of one study can accurately predict the AD status of the patients in another. We find greater than 90\% recall when training on ADNI1 or ADNI2 and testing on the other two studies. However, when the model is trained on the ADNIGO dataset, the recall when testing on ADNI1 or ANDI2 plummets to approximately 14\%. We expect this is due to the small number of Alzheimer's patients in the ADNIGO dataset and the small sample size. Ultimately, this analysis shows that our best models generalize well on different training and testing subsets of ADNI data, as long as there is a sufficiently large training set size with a moderate amount of Alzheimer's patients. 

\subsection*{Validation of Model with AIBL Data}

One of the stretch goals for this study was the evaluation of our predictive models with patients that are part of a completely different repository such as AIBL. To this end, we considered a total of 861 patients from the AIBL database (note: for the ADNI repository we considered 1737 patients). Since AIBL uses slightly different protocols, a direct merge of both databases was not possible. Thus, we focused our analysis on a limited number of features that are consistently available in both ADNI and AIBL patients. Although small in size, this set of features includes the demographics of the patients (age and gender), their genetic characterization (APOE4) and their responses to a well-known cognitive test (MMSE). An important consideration is that the distribution of these features across the ADNI and AIBL populations is not exactly the same. The ADNI dataset has more men than women, and AIBL patients tend to be older than ADNI patients. The distributions of MMSE scores and APOE4 results look approximately the same between the two studies, after taking into account the difference in samples sizes. 

We conducted two different studies involving the AIBL data:
\begin{itemize}
\item AIBL Test 1: We merged all patients (i.e., ADNI and AIBL) into a single set and then split them (randomly) into training and test sets (without consideration of what repository patients belong to). From here we ran 100 different simulations using Model 1 (Random Forest with mean/mode imputation, no feature selection, and 1,000 trees).
\item AIBL Test 2: In this study we trained a model using all ADNI patients and then we assigned all AIBL patients to the test set. We ran 100 different simulations using Model 1.
\end{itemize}

Results for accuracy, recall, prediction and f1 score for these two tests using AIBL data are summarized in Table~\ref{tab:AIBLModelComparisonTable}. These results demonstrate that when merging ADNI and AIBL patients into one single study our predictive model still yields high accuracy (90.0\%) and recall (85.0\%). When training on ADNI patients and testing on AIBL patients, 
we also obtain high accuracy; however, we observe a lower recall and precision. We expect that these results could be improved with domain adaptation and the introduction of more common features to both AIBL and ADNI.

 \begin{table}[htb]
\centering
\caption{Metrics for models that consider AIBL data}
\label{tab:AIBLModelComparisonTable}
\begin{tabular}{|c|c|c|c|c|}
\hline
                 & \textbf{Accuracy} & \textbf{Recall} & \textbf{Precision} & \textbf{F1 Score} \\ \hline
\textbf{AIBL Test 1} & 90\%             & 85\%           & 82\%              & 83\%             \\ \hline
\textbf{AIBL Test 2} & 93\%             & 80\%           & 69\%              & 74\%             \\ \hline
\end{tabular}
\end{table}

\subsection*{Longitudinal Data Analysis}
\begin{figure}[htb]
\centering
\includegraphics[width=0.7\textwidth]{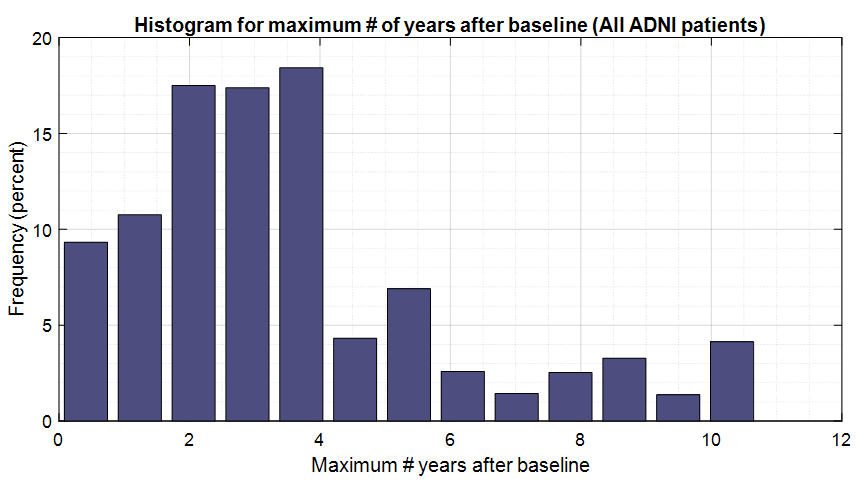}
\caption{\label{fig:Distribution_Years} Histogram for maximum number of years ADNI patients have been studied}
\end{figure}

Next, we investigated how the predictive power of our best models are affected by the number of visits available for each patient. Number of visits is strongly correlated with the length of time a patient has been in the study. We also investigate the distribution of the maximum number of years that ADNI patients have been studied (see Figure~\ref{fig:Distribution_Years}). This figure shows that approximately 80\% of the ADNI patients have endured 4 years or less of study under this protocol, while only 20\% of the patients have undergone more than 4 years of study.

Using Model 1 (as used in Table~\ref{tab:ModelComparisonTable}) we ran 100 samples for random splits of the ADNI patients into training and validation sets. For the training set we assumed that the full history of visits was known (history is known). However, for patients in the test set, we fixed the maximum number of visits that could be used in computing the longitudinal features of the patients. Results, which are shown in Table~\ref{tab:LongitudinalTable}, clearly indicate that the performance of the predictive model improves as the number of visits for test patients increase. This trend is expected, as additional visits provide more valuable information about the patient evolution. The main contribution of Table~\ref{tab:LongitudinalTable} is the quantification of this trend. We see, for example, that achieving 89\% accuracy requires at least 6 clinical visits for an average patient (meaning approximately 2.5 to 4.5 years of study). A guarantee of 85\% recall on the prediction would require around 16 visits for the average patients (roughly equivalent to 8-10 years of study). Hence this table potentially provides a valuable tool for doctors and patients to understand the number of visits required in order to obtain predictions for AD with a satisfying confidence level.


\begin{table}[htb]
\centering
\caption{Longitudinal Analysis: prediction metrics vs. number of visits}
\label{tab:LongitudinalTable}
\begin{adjustbox}{width=1\textwidth}
\begin{tabular}{|c|c|c|c|c|c|c|}
\hline
                \textbf{\# Visits} & \textbf{\# Years since baseline [range]} & \textbf{\# Test Patients} & \textbf{Accuracy} & \textbf{Recall} & \textbf{Precision} & \textbf{F1 Score} \\ \hline
\textbf{1} & [0.3 - 1] & 574 & 79             & 65           & 79              & 70             \\ \hline
\textbf{2} & [0.5 - 2]  & 550 & 86             & 75           & 88              & 81             \\ \hline
\textbf{4} & [1.5 - 3]  & 494 & 87             & 78           & 88              & 83             \\ \hline
\textbf{6} & [2.5 - 4.5]  & 371 & 89             & 81           & 86              & 83             \\ \hline
\textbf{8} & [4 - 6]  & 221 & 89             & 83           & 84              & 84             \\ \hline
\textbf{12} & [6 - 8]  & 81 & 89             & 80           & 89              & 84             \\ \hline
\textbf{16} & [8 - 10]  & 45 & 90             & 85           & 89              & 87             \\ \hline
\end{tabular}
\end{adjustbox}
\end{table}

\subsection*{Cost Analysis and Meta-Classification}
\begin{figure}
\centering
\includegraphics[width=0.7\textwidth]{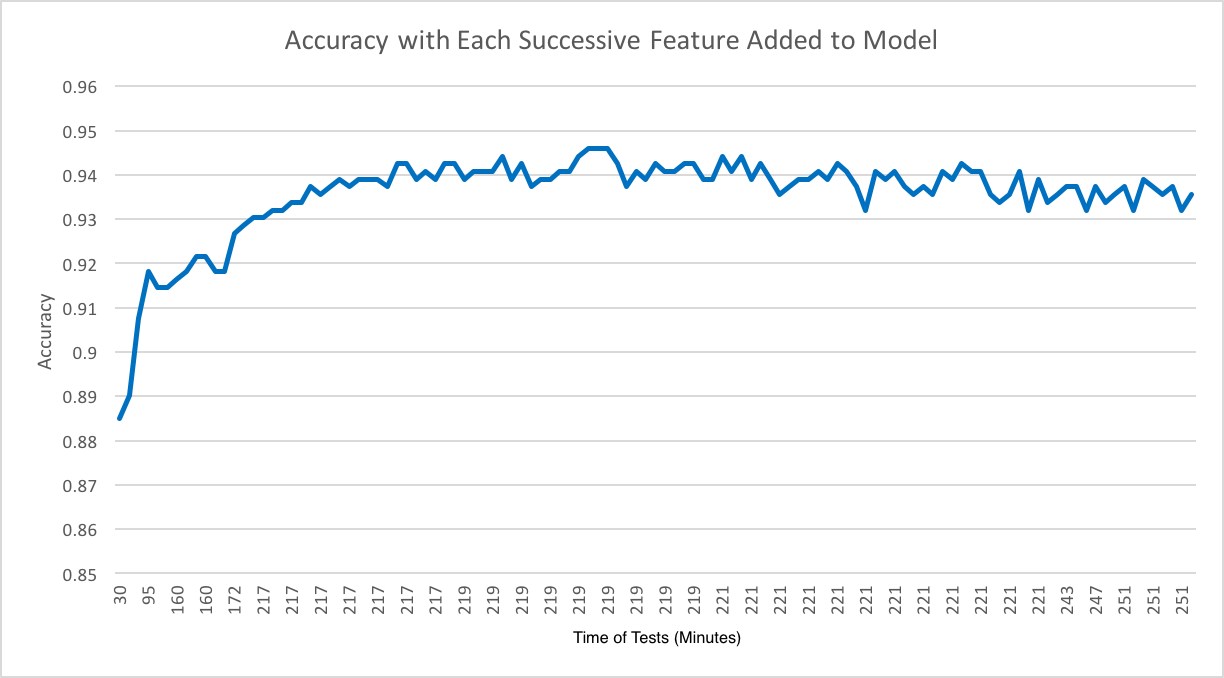}
\caption{\label{fig:featurecost} Accuracy with Successive Features Added to Model versus Time of Medical Tests}
\end{figure}

When we plot the time needed to obtain the features for each successive model versus the accuracy of these models, we see that after approximately $217$ minutes of testing, the accuracy of the models plateau (Figure \ref{fig:featurecost}). Note that when a longitudinal metric, for example mean CDRSB, is added to the model already containing another longitudinal metric for the same feature, e.g. standard deviation of CDRSB, no extra testing time is required to include this feature. Therefore, the x-axis contains duplicated values.  

\begin{figure}
\centering
\includegraphics[width=\textwidth]{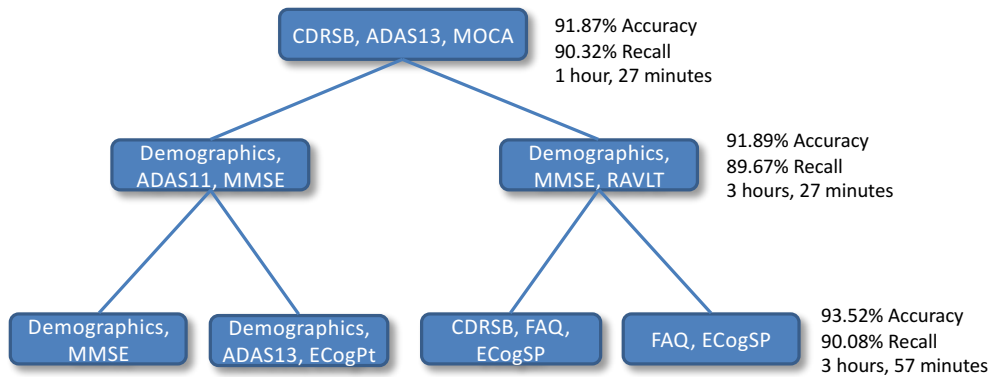}
\caption{\label{fig:meta} Example of Decision Tree produced by Meta-Classification}
\end{figure}

An example of a meta-classification Decision Tree is included in Figure \ref{fig:meta}, with the accuracy and recall scores averaged over fifty random training and test set splits to ensure robustness of results. The accuracy, recall and time needed to perform the tests are provided for each level in the tree. Note that accuracy increases by level, but that recall peaks at Level 1, most likely because this is a more generalizable model. Note that even with the the Level 1 tree, trained on just three cognitive tests, CDRSB, ADAS13, and MOCA, which only take a combined $1$ hour and $27$ minutes, our model is able to predict with better than 90\% accuracy and recall. The CDRSB (Clinical Dementia Rating Box Score) takes $30$ minutes approximately and is scored based on the results from an interview with the patient and the patient's caregiver. ADAS11 is one of the most popular cognitive tests for AD consisting of a $45$ minute written test containing $11$ questions. Finally MOCA (Montreal Cognitive Assessment) is a brief written test that takes approximately $12$ minutes to complete. Note that CDRSB and ADAS11 rank in the top five most important features  in Figure \ref{fig:featimp}. The inclusion of MOCA is most likely a result of its low time cost.  \\

While we currently only consider the cost of features in terms of the patient's time, this model could easily incorporate the monetary cost of these tests as well. These results show that this type of meta-classification model can perform very well, suggesting its possible implementation as a data-driven diagnostic tool. Using this model, the patient and doctor can weigh whether the added specificity and sensitivity warrant the extra time and cost of the medical test. 

We also considered the possibility of combining longitudinal analysis with meta-classification. For this analysis, we took the features in the Level 1 model (CDRSB, ADAS13 and MOCA) and ran an analysis similar to the one presented in the Longitudinal Analysis section. Results are presented in Table~\ref{tab:LongitudinalTable2}. Once again, this table quantifies the performance of the predictive models (in this case using only the 3 features of Level 1 tree) for different numbers of clinical visits. Compared to Table~\ref{tab:LongitudinalTable}, the number of visits required to achieve 89\% accuracy using only these three features would be larger (16 vs 6). Nevertheless, these results indicate that good performance for AD prediction can still be achieved using a limited number of visits and a reduced set of features.


\begin{table}[htb]
\centering
\caption{Longitudinal Analysis: prediction metrics vs. number of visits using Level 1 tree features from meta-classification analysis}
\label{tab:LongitudinalTable2}
\begin{adjustbox}{width=1\textwidth}
\begin{tabular}{|c|c|c|c|c|c|c|}
\hline
                \textbf{\# Visits} & \textbf{\# Years since baseline [range]} & \textbf{\# Test Patients} & \textbf{Accuracy} & \textbf{Recall} & \textbf{Precision} & \textbf{F1 Score} \\ \hline
\textbf{1} & [0.3 - 1] & 574 & 78             & 64           & 77              & 69             \\ \hline
\textbf{2} & [0.5 - 2]  & 550 & 85             & 77           & 84              & 81             \\ \hline
\textbf{4} & [1.5 - 3]  & 494 & 87             & 79           & 86              & 82             \\ \hline
\textbf{6} & [2.5 - 4.5]  & 371 & 87             & 81           & 79              & 81             \\ \hline
\textbf{8} & [4 - 6]  & 221 & 87             & 84           & 80              & 82             \\ \hline
\textbf{12} & [6 - 8]  & 81 & 87             & 80           & 87              & 82             \\ \hline
\textbf{16} & [8 - 10]  & 45 & 89             & 87           & 87              & 86             \\ \hline
\end{tabular}
\end{adjustbox}
\end{table}

\section*{Discussion}
Our work shows that it is possible to build a data-driven model that can confidently predict the risk of developing Alzheimer's in the future with a level of accuracy and recall that are above 90\%. The necessary data for such a prediction is patient demographic information, a genetic test (APOE4 genotyping) and a battery of cognitive tests. We demonstrated that imaging data (MRI and PET scans), which are more costly in terms of time and money, are not necessary for highly accurate predictions. We also demonstrated how well our model generalizes by evaluating the model performance for different ADNI sub-studies (testing one against the others and quantifying model performance) and against a cohort of patients that belong to a completely different repository (AIBL). In all cases, our predictive models show very robust performance.\\  

We carefully quantified the impact that the number of clinical visits of data available for a patient has on the predictive performance of our model. We also implemented a meta-classification technique to identify the combination of features that provide the optimal balance between model prediction and feature cost. In each case we have identified models that can still provide a high level of accuracy and recall. We believe our work provides the right framework for a practical deployment of an AD predictive tool in clinical settings. As an example, we have proposed a diagnostic protocol with only 3 tests and 4 clinical visits that can predict AD with 87\% accuracy and 79\% recall. Ultimately our model framework could be used by physicians and patients together to determine appropriate plans for diagnosis and monitoring of the risk of developing AD. \\

Any potential model to be deployed in real world settings will have to perform well relative to a clinician. Based on the literature, physicians can diagnose Alzheimer's with 87\% accuracy and 91\% recall \cite{doctor}. Our best models produce equivalent or better predictions relative to physicians for the harder problem of predicting \underline{future} development of AD. Going forward, a parallel study of model prediction versus physician prediction would be necessary to validate the models and gain doctor’s trust in this method. A limitation to this approach is due to our training labels being provided by doctors. Those "true" labels carry some level of uncertainty as AD is a difficult disease to diagnose in vivo. Our predictive models are ultimately only as good as the training data used to build them. Finally, it is important to recognize that this work has focused on proposing models that offer high predictive performance, with no consideration for interpretation of these models. Expected FDA new regulations for CDS (Clinical Decision Support) software could incentivize developing models. \\ 

\section*{Acknowledgments}
Data used in preparation of this article were obtained from the Alzheimer's Disease Neuroimaging Initiative (ADNI) database (adni.loni.usc.edu). As such, the investigators within the ADNI contributed to the design and implementation of ADNI and/or provided data but did not participate in analysis or writing of this report. A complete listing of ADNI investigators can be found at http://adni.loni.usc.edu/wp-content/uploads/how\_to\_apply/ADNI\_Acknowledgement\_List.pdf


%
%
%


\end{document}